# Are European equity markets efficient?

# New evidence from fractal analysis

Enrico Onali[1,2]

John Goddard[1]

**Abstract:** Fractal analysis is carried out on the stock market indices of seven European countries and the US. We find evidence of long-range dependence in the log return series of the Mibtel (Italy) and the PX-Glob (Czech Republic). Long-range dependence implies that predictable patterns in the log returns do not dissipate quickly, and may therefore produce potential arbitrage opportunities. Therefore, these results are in contravention of the Efficient Market Hypothesis. We show that correcting for short-range dependence, or 'pre-filtering', may dispose of genuine long-range dependence, suggesting that the market is efficient in cases when it is not. Pre-filtering does not reduce significantly the power of the tests only for cases for which the Hurst exponent (a measure of the long-range dependence) lies well outside the boundaries of no long-range dependence. For borderline cases, the pre-filtering procedure reduces the power of the test. On the other hand, the absence of pre-filtering does not result in a test that is significantly over-sized.

[1] Bangor Business School, Bangor University, Bangor, Gwynedd, LL57 2 DG, UK

[2] Corresponding author: e.onali@bangor.ac.uk

# 1 Introduction

Traditional capital markets theory relies on the assumption that log prices are martingales, implying that the expected value of log price is the log price in the previous period and log returns are temporally independent. Log prices follow random walks and are unpredictable. The Random Walk Hypothesis (RWH) is a variant of the Efficient Market Hypothesis (EMH). Temporal dependence in log returns is inconsistent with the EMH, and with mean-variance portfolio theory (Markovitz, 1952, 1959) and the capital asset pricing model (CAPM). While short-range dependence is unlikely to lead to systematic abnormal returns, long-range dependence, under certain conditions, implies that trading strategies based on historical prices may be systematically profitable (Mandelbrot, 1971; Rogers, 1997). The existence of arbitrage opportunities is contrary to the RWH and the EMH.

In order to allow for dependence in log returns and other stylized facts about stock market behaviour, Peters (1994) introduces the Fractal Markets Hypothesis (FMH). The FMH does not reject a priori the assumption that returns are independent, but allows for a broader range of behaviour. As a result, the FMH does not necessarily constitute an alternative to the EMH, but rather a generalisation. The FMH derives from the theory of fractals (Mandelbrot, 1982). A fractal is an object whose parts resemble the whole. Peters (1994) argues that markets have a fractal nature: when markets are stable, returns calculated over different time scales (daily, weekly, monthly, and so on) exhibit the same autocovariance structure. For instance, if daily returns exhibit positive dependence, so too do monthly returns. This feature is called self-affinity.

The autocovariance structure of a self-affine time series can be represented by the Hurst exponent. If the variance and higher-order moments of the returns series are finite, a Hurst exponent of 0.5 indicates that log returns are random. A Hurst exponent larger (smaller) than 0.5 indicates positive (negative) long-range dependence in the returns series calculated over any time scale. For positive long-range dependence in the log returns, the autocorrelation function for log returns calculated over any time scale decays slowly. If the variance and higher-order moments of the returns series are infinite, the Hurst exponent may take a value larger than 0.5, even when returns are random and there is no long-range dependence. The Rescaled Range Analysis (RRA) has been employed widely to calculate the Hurst exponent, interpreted as an indicator of long-range dependence. The RRA is robust to departures from Normality, including the cases where the variance and higher-order moments are either finite or infinite.

Empirical studies on the long-range dependence properties of returns precede the FMH (Peters, 1994). Initially, evidence of long-range dependence is found for the US stock market (Greene and Fielitz, 1977; Peters 1991). Subsequent refinements of the methodology used to measure long-range dependence have produced results consistent with the EMH (Lo, 1991). Recently, Serletis and Rosenberg (2009) fail to find evidence of long-range dependence for four US stock market indices. International equity markets have also been examined (Cheung and Lai, 1995; Jacobsen, 1996; Opong *et al.*, 1999; Howe *et al.*, 1999; McKenzie, 2001; Costa and Vasconcelos, 2003; Kim and Yoon, 2004; Zhuang *et al.*, 2004; Norouzzadeh and Jafari, 2005; Onali and Goddard, 2009), as well as commodities markets (Cheung and Lai, 1993; Alvarez-Ramirez *et al.*, 2002; Serletis and Rosenberg, 2007), and exchange rates (Mulligan, 2000; Kim and Yoon, 2004; Da Silva *et al.*, 2007). Recently, evidence has been reported of a connection between the Hurst exponent and market crashes (Grech and Mazur, 2004; Grech and Pamula, 2008). Cajueiro and Tabak (2004, 2005) rank the efficiency of emerging markets using estimated Hurst exponents for either stock returns and volatility. The higher the Hurst exponent, the lower is the efficiency of the stock market.

There is still lack of consensus regarding the method to use for estimating the degree of long-range dependence. Moreover, many studies base their conclusions on the estimated Hurst exponent in the absence of a sound methodology for hypothesis testing. Any inference drawn on the basis of estimated Hurst exponents without a comparison with critical values generated using standard hypothesis testing criteria is questionable.

This paper presents tests for the validity of the EMH for eight stock markets that are believed to be at different stages of development. We examine stock market indices comprising a large number of stocks. Predictability due to thin trading is therefore unlikely. Rejection of the hypothesis of long-range independence in favour of long-range dependence would suggest the possible existence of arbitrage opportunities.

We contribute to the extant literature in several ways. First, we provide evidence of significant long-range dependence for the PX-Glob, the index of the Czech stock market, and for the Mibtel, the index of the Italian stock market. Significant long-range dependence is contrary to the EMH, because it implies that dependence in the log returns does not dissipate quickly, and could give rise to arbitrage opportunities. For the other six market indices, there is no evidence of significant long-range dependence. However, there is evidence of short-range dependence (except for the US stock market), and non-Normality in returns calculated at high frequencies, contrary to the assumptions of portfolio theory and CAPM.

Second, unlike most of the extant literature, we employ Monte Carlo simulations to construct critical values for the null hypothesis of independent and Normal log returns. This enables us to assess whether the estimated Hurst exponents for the log return series are long-range dependent, contrary to the EMH.

Third, previous literature uses the RRA on the residuals of an autoregressive model of log returns, to prevent any short-range dependence in the data from producing spurious detection of long-range dependence or persistence (Peters, 1994; Jacobsen, 1996; Opong *et* *al.*, 1999). Other methods that control for short-range dependence are suggested by Lo (1991), and Fillol and Tripier (2004). We present Monte Carlo simulation evidence to suggest that pre-filtering to remove short-range dependence may impair the self-affine structure of the returns series, and the capability of the RRA to detect long-range dependence. We run the RRA both with and without pre-filtering, and compare the results. The RRA applied to the filtered log returns fails to reject the RWH for any of the indices, except for the PX-Glob. However, when the RRA is run on the unfiltered returns series, evidence contrary to the RWH is found even for the Mibtel, for which the Hurst exponent is significantly larger than 0.5. Monte Carlo simulations using return series with the same short-range dependence structure as the log returns of Mibtel and either long-range dependence or long-range independence show that for low values of the Hurst exponent pre-filtering removes genuine long-range dependence. Therefore, we posit that the choice to pre-filter a log return series for cases for which there is long-range dependence but the strength of the dependence is modest may be critical for the correct detection of long-range dependence. For 'borderline' cases, pre-filtering may dispose of genuine long-range dependence, while an examination of the original log return series is unlikely to conduce to a test that is over-sized.

The rest of the paper is structured as follows. Section 2 reviews the properties of self-affinity, long-range dependence, and the generalised Central Limit Theorem. Section 3 describes the methodology and data. Section 4 reports the results. Section 5 concludes.

## 2  Self-affinity and long-range dependence

Self-similarity is the distinguishing feature of fractals: each part comprising a fractal resembles the whole. In a financial asset log returns series, the weaker concept of self-affinity is employed: after rescaling using a factor that depends on the scale only, a self-affine returns series has the same distributional properties irrespective of the scale over which the returns are measured (daily, weekly, monthly, and so on).

Self-affinity is described as follows (Calvet and Fisher, 2002, p. 383):

$$\{X(nt_1),...,X(nt_k)\} \stackrel{d}{=} \{n^H X(t_1),...,n^H X(t_k)\} \tag{1}$$

where $X(t)$ denotes log returns measured over period t, $H > 0$, $n$, $k$, $t_1...t_k \geq 0$, and $\stackrel{d}{=}$ denotes equality in distribution.

Self-affinity in a log returns series with Normal increments implies that the variance of log returns, $\gamma_0^{(n)}$, scales proportionately with the time scale over which the returns are measured, $n$, according to a factor of proportionality governed by the Hurst exponent, $H$:

$$\gamma_0^{(n)} = n^{2H} \gamma_0^{(1)} \tag{2}$$

Self-affinity implies that the autocorrelation function for lag $k$, $\rho_k^{(n)} = \gamma_k^{(n)} / \gamma_o^{(n)}$, is the same for all n, or $\rho_k^{(n)} = \rho_k^{(1)}$ for n $\geq$ 1 and k $\geq$ 1. In this case, $\gamma_k^{(n)} = (1/2)[(k+1)^{2H} - 2k^{2H} + |k-1|^{2H}]$. A fractionally-integrated log returns series, which gives rise to a log price series characterised as Fractional Brownian Motion (FBM), satisfies the property of self-affinity. FBM is a generalisation of Brownian Motion, the continuous-time analogue of the random walk. FBM has similar features to Brownian Motion, but with increments that are long-range dependent and therefore non-random (Mandelbrot and van Ness, 1968). Long-range dependence (at all time scales) in a self-affine series is represented by the parameter 0 < $H$ < 1. For 0 < $H$ < 0.5, values of the series $k$ periods apart are negatively correlated for all $k$, or antipersistent. For 0.5 < $H$ < 1, values of the series $k$ periods apart are positively correlated or persistent. Self-affinity implies that '[…] the distribution of returns over different sampling intervals is identical except for a single, non-random contraction' (Mandelbrot, Fisher and Calvet, 1997, p. 8). If a log returns series is either persistent ($H$ > 0.5) or antipersistent ($H$ < 0.5), the Efficient Market Hypothesis (EMH) is violated. An extensive literature examines whether the EMH correctly represents the behaviour of stock market returns, using fractal analysis. If the EMH is rejected, the FMH may represent a better explanation for the behaviour of stock returns. The Rescaled Range Analysis (RRA) is a widely used method for estimation of long-range dependence, which is robust to non-Normality in the log returns (Mandelbrot and Taqqu, 1979).

The best-known discrete-time representation of long-range dependence is provided by the Autoregressive Fractionally Integrated Moving Average (ARFIMA) model (Granger and Joyeux, 1980; Hosking, 1981). The ARFIMA(0,$d$,0) process, where $d = H - 0.5$, is asymptotically self-affine (Fisher, Calvet and Mandelbrot, 1997). The ARFIMA($p$,$d$,$q$) process accommodates both short-range and long-range dependence. Due to the presence of short-range dependence, estimation of the Hurst exponent using RRA may be upward biased. Lo (1991) adjusts the original RRA to allow for short-range dependence, by including weighted autocovariances for up to $l$ lags in the log returns. Teverovsky et al. (1999) present a Monte Carlo analysis to show that the larger is $l$, the lower is the power of the test, and the higher is the probability that the null hypothesis of absence of long-range dependence is accepted when it is false. Andrews (1991) suggests a procedure for selecting the number of autocovariances to include in the calculation of the adjusted RRA, based on the first-order autocorrelation coefficient. Jacobsen (1996) runs the RRA on the residuals of an AR(1) and an MA(1) model. Fillol and Tripier (2004) use Generalised Least Square (GLS) to estimate the Hurst exponent, using the scaling function (Mandelbrot, Calvet, and Fisher, 1997).

None of these studies provides a clear-cut solution for discriminating between short-range and long-range dependence. Jacobsen (1996) finds that filtering the original returns using short-range dependent models reduces the power of the test when the true process is long-range dependent, or when the process contains both short-range and long-range dependence. Fillol and Tripier (2004) report similar findings.

Table 1 examines the impact of pre-filtering on the RRA estimate of the Hurst exponent when the true data generating process for log returns is characterised by zero dependence, ARFIMA(0,0,0); by long-range dependence only, ARFIMA(0,$d$,0); by short-range dependence only, ARFIMA(1,0,0); and by both long-range and short-range dependence, ARFIMA (1,$d$,0). We consider $T = 1000$ and the following permutations of values for $\rho$ (the first-order autoregressive parameter) and $d$ (the order of fractional integration): $\rho = 0, 0.10, 0.20$ and $d = 0, 0.04, 0.08, 0.12$.[1]

The mean estimated Hurst exponent in the case of ARFIMA(0,0,0) is $\hat{H} = 0.6130$. The true value is $H = 0.5$ and the estimated Hurst exponent is upward biased, as is well known. After pre-filtering ARFIMA(0,0,0) by fitting an AR(1) model and applying the RRA to the residuals, the mean estimated Hurst exponent is virtually unaltered, $\hat{H} = 0.6129$. The critical values for the rejection of the null hypothesis of zero long-range dependence are 0.6385, 0.6459 and 0.6631 at the 0.1, 0.05 and 0.01 significance levels, respectively. Short-range dependence imparts additional upward bias into the estimated Hurst exponent, resulting in a test that is over-sized. For ARFIMA(1,0,0), the mean estimated Hurst exponent is $\hat{H} = 0.6388$ for $\rho = 0.10$ and $\hat{H} = 0.6654$ for $\rho = 0.20$. Pre-filtering ARFIMA(1,0,0) eliminates the spurious detection of long-range dependence. However, pre-filtering ARFIMA(1,$d$,0) may produce a bias in the opposite direction. For $d = 0.08$ and $\rho = 0.20$, the mean estimated Hurst exponent for the unfiltered series is $\hat{H} = 0.7053$, while the mean estimated Hurst exponent for the pre-filtered series is $\hat{H} = 0.6280$. The latter value is less than one standard deviation from the mean estimated Hurst exponent for ARFIMA(0,0,0). Pre-filtering therefore reduces the power of the test for long-range dependence. Similarly for $d = 0.12$ and $\rho = 0.20$, the mean estimated Hurst exponent for the unfiltered series is $\hat{H} = 0.7249$, while the mean estimated Hurst exponent for the pre-filtered series is 0.6300.

---

[1] In an evaluation of the ability of RRA to discriminate between short-range and long-range dependence, Jacobsen (1996) considers $\rho = 0.268, -0.230$ and $d = 0.10, 0.20, 0.30$. These selected values for $\rho$ appear unrealistically large in absolute terms for log returns series in developed stock markets.

[insert Table 1 here]

## 3   Methodology and data

In this paper, we use the Rescaled Range Analysis (RRA) to estimate the Hurst exponent and to assess the degree of long-range dependence for eight stock market indices. Section 3.1 describes the estimation method. Section 3.2 reports descriptive statistics for the returns series for the eight market indices.

*3.1   Methodology*

The RRA is based on an examination of the average rescaled range of the cumulative deviation of a time series from its mean value within each of a number of subperiods. The rescaled range statistic, denoted $(R/S)_n$, is specific to the time scale, $n$, equivalent to the number of daily returns observations included in each subperiod. To obtain an indication of the scaling behaviour of $(R/S)_n$ as $n$ varies, $(R/S)_n$ is constructed so as to vary proportionately with $n^H$.

Let $N$ denote the total number of observations in the series $z_t$, where $z_t$ represents either the residuals from an AR($p$) model fitted to the log returns series[2] in the case of pre-filtering, or the unfiltered log returns series. Starting from the first observation, subdivide $N$ into $M$ contiguous subperiods labelled $m = 1,...,M$, each containing $n$ observations such that $N - n < Mn \leq N$. For the observations within subperiod $m$, the mean and standard deviation of $z_t$ are:

---

[2] The number of lags of the AR model, $p$, is equal to the lags for which the Partial Autocorrelation Function (PACF) is statistically significant at the 5%. Only lags up to the tenth lag are considered.

$$\mu_m = n^{-1} \sum_{t=(m-1)n+1}^{mn} z_t \tag{4}$$

$$S_m = \sqrt{n^{-1} \sum_{t=(m-1)n+1}^{mn} (z_t - \mu_m)^2} \tag{5}$$

for $m = 1,...,M$.

The cumulative deviations of $z_t$ from $\mu_m$ within subperiod $m$ are:

$$x_t = \sum_{s=(m-1)n+1}^{t} (z_s - \mu_m) \tag{6}$$

where $t = (m-1)n+1,...,mn-1$, and $x_{nm} = 0$.

The range for subperiod $m$ is defined as the difference between the maximum and minimum values of $x_t$ for the observations within subperiod $m$:

$$R_m = \max_{t \in m}(x_t) - \min_{t \in m}(x_t) \tag{7}$$

Commonly, $N$ is not a multiple of $n$. If $Mn < N$, then $L = N - nM$ observations at the end of the observation period are unused in the above procedure. To avoid discarding these $L$ observations, the procedure is repeated starting from the $L+1$th observation (rather than from the first observation). A second set of $M$ calculated values of $R_m$ and $S_m$ is obtained, where $m=M+1,...,2M$. If $Mn = N$, $R_m$ and $S_m$ for $m = 1,...,M$ are identical to $R_m$ and $S_m$ for $m = M+1,...,2M$.

The $(R/S)_n$ statistic is the mean of the rescaled range values for $m = 1,...,2M$

$$(R/S)_n = (2M)^{-1} \sum_{m=1}^{2M} (R_m / S_m) \tag{8}$$

Finally, the scaling behaviour of $(R/S)_n$ can be investigated by examining the power-law relationship $(R/S)_n \sim n^H$, where $H$ is the Hurst Exponent. Having obtained values of $(R/S)_n$ for several time scales $n$, $H$ can be estimated by running the ordinary least squares (OLS) regression

$$\ln[(R/S)_n] = \ln(c) + H \ln(n) + \eta_n \tag{9}$$

where $\eta_n$ is a disturbance term.

As noted above, the RRA is able to identify long-range dependence regardless whether the random component of the log returns series is Gaussian or not (Mandelbrot and Taqqu, 1979). Several previous studies have noted that the application of the RRA in finite samples produces an upward-biased estimator of $H$ (Feller, 1951; Anis and Lloyd, 1976; Peters, 1994; Qian and Rasheed, 2004; and Norouzzadeh and Jafari, 2005). Following Onali and Goddard (2009), Monte Carlo simulations are used to examine whether the estimated Hurst exponents for the log return series of the indices differ significantly from the value that is expected under the EMH. This procedure involves generating 5,000 simulated returns series containing random Normal innovations. The RRA is calculated for each of the simulated series, and the sampling distributions are obtained for the estimated $H$.

Table 2 examines the power functions of tests for $H_0$: $H = 0.5$ against $H_1$: $H > 0.5$ at the 0.05 significance level based on the RRA estimator, and compares with tests based on the scaling function method developed by Mandelbrot, Calvet, and Fisher (1997), when the true generating process is ARFIMA(0,$d$,0) with $d = H - 0.5$. We consider sample sizes of $T = 1000, 2000, 5000, 10000$. In general, the test based on the RRA has higher power than the test based on the scaling function. These results provide the justification for our preference for the test based on the RRA.

[insert Table 2 here]

### 3.2 Data and descriptive statistics

Daily log returns are calculated, based on closing daily prices for the period 31/08/1995-30/08/2005 provided by Thomson Analytics, for eight stock market indices. Six of these stock markets are located in Western Europe: Mibtel (Milan), CDAX (Frankfurt), FTSE 350 (London), Amsterdam S.E. all-share (henceforth, ASE), Madrid S.E. all-share (MSE), Swiss Market Index (SWX, Zurich). As a term of comparison, we also examine an index from an Eastern European country, the PX-Glob (Prague), and an index from the US stock market, the Dow Jones Industrial Average (New York Stock Exchange, hencefort DJIA). The Czech Republic was undergoing a period of economic transition in the period under investigation, and its stock market was relatively under-developed.[3] The US stock market is widely accepted as the most efficient stock market worldwide. As a result of the posited link between market development, market efficiency, and the degree of long-range dependence in returns, we expect the PX-Glob to exhibit long-range dependence,[4] while the DJIA should not exhibit long-range dependence. For the six Western European countries, we expect either no long-range dependence, or a modest long-range dependence ($H$ slightly above 0.5 or $d$ slightly above 0). As a first indication of the level of dependence in a developing stock market, the first-order autocorrelation for PX-Glob is 0.1176, much larger than for the other six European stock markets (for which $\rho_1$ ranges between -0.0026, for the CDAX, and 0.0351, for the SWX). The first order autocorrelation for the DJIA is barely -0.0106 and the PACF for lags from 1 to 10 is never significant.

Table 3 reports descriptive statistics for the log returns series for four time scales: daily, weekly, monthly, and quarterly.[5] The total observations for the four series are 2,608, 522, 120, and 40, respectively. The descriptive statistics shown in Table 3 relate to the first four central moments of the distribution of the log returns: mean, standard deviation, skewness, and kurtosis.

[insert Table 3 here]

---

[3] For a report on the economic reforms in the Czech Republic throughout the 1990s, see OECD (2001).
[4] Because of missing observations over the period 31/08/1995-30/08/2005 for the PX-Glob, we extend the sample period for PX-GLOB until 01/02/2006. This extension enables us to have 2608 return observations, as for all other countries.
[5] Given $n$ the number of trading days comprising each time scale, for the daily returns, $n = 1$. For the other time scales, there is no exact correspondence between $n$ and the actual time scale used in the analysis, as the number of trading days may vary according to the week, month, or quarter.

The skewness is negative in most cases. The index with the highest level of skewness (in absolute terms) for the daily frequency is the PX-Glob, and the index with the lowest level of skewness for the daily frequency is the SWX. These results may suggest that in developed markets large negative returns (market crashes) are more likely than in developed markets. As the time scale increases, the distributions of the log returns series do not appear to become more symmetric. The departure from Normality for quarterly data is contrary to the alleged phenomenon of 'aggregational Normality' (Cont, 2001). The presence of negative skewness for long time scales suggests that negative daily returns tend to cluster, making large cumulative losses more likely than large cumulative gains. Negative skewness implies large losses are more likely than large gains. *Ceteris paribus*, negative skewness should encourage investors to require higher expected returns than if the skewness is zero.

The index with the highest level of kurtosis for the daily frequency is the DJIA, and the index with the lowest level of kurtosis for the daily frequency is the PX-Glob. These results may suggest that in developed markets the information flow is faster, and news are incorporated quickly in prices, causing large daily price fluctuations. The distributions of the log returns series seem to become less leptokurtic as the time scale increases (although for the FTSE 350 the quarterly log returns are more leptokurtic than the monthly log returns). Leptokurtosis implies extreme returns of either sign are more likely than in a Normal distribution. *Ceteris paribus,* investors should require higher expected returns when the distribution is leptokurtic than in the case of a mesokurtic distribution (where the kurtosis is three).

Provided the population variance is finite, the Central Limit Theorem (CLT) ensures that the sum of independent random variables converges to a Normal distribution as the number of variables increases, regardless of the distribution of each individual random variable. If the variance of the population is infinite, however, the CLT does not hold. In this case, the sum of IID variables converges to a Stable Paretian distribution (Levy, 1925) with a characteristic exponent equal to that of the individual random variables. Independent Stable Paretian variables satisfy the property of 'stability under addition', which is a generalisation of the Central Limit Theorem (CLT). In other words, the Stable Paretian distribution is able to account for self-affinity of a time series with independent non-Normal increments. If log returns converge to a Stable Paretian distribution with infinite variance, many concepts related to the EMH are invalid.

In order to investigate whether the log returns series comply with the CLT once temporal dependence has been eliminated, the procedures described above are repeated on shuffled surrogates of the log return series, in which the ordering of the observations is randomised.

Table 4 reports descriptive statistics for the shuffled surrogate of the log return series for four time scales: daily, weekly, monthly, and quarterly. There is little evidence of leptokurtosis for long time scales. For the PX-Glob, the kurtosis is lower than for a Normal distribution (*platokurtosis*) for the quarterly time scale. In comparison with the results reported in Table 3, the degree of skewness is also less for long time scales. The reduction in skewness for long time scales in the shuffled series supports the hypothesis that negative skewness for long time scales may be due to clustering of negative daily returns. These results are consistent with the CLT.

## 4     Results

The Rescaled Range Analysis (RRA) is carried out over 40 values of the time scale parameter $n$, defined by increasing $\ln(n)$ in steps of 0.1 from a minimum of $\ln(n) = 1.6$ ($n = 5$) to a maximum of $\ln(n) = 5.7$ ($n = 299$). For each value of $\ln(n)$, $n$ is obtained by rounding $e^{[\ln(n)]}$ to the nearest integer. Two of the 42 values of ln(*n*) in the range (1.6, 5.7) are discarded, because the integer values of $e^{[\ln(n)]}$ and $e^{[\ln(n-0.1)]}$ are identical. The results of the RRA for each index are compared with critical values obtained via Monte Carlo simulations. To investigate whether there is any statistical evidence of departure from the RWH (due to either persistence or antipersistence) when returns are measured over small or large time scales only, the estimation of equation (9), and the Monte Carlo simulations, are repeated, by fitting a spline function to allow for a change in the estimated Hurst exponent.

Tables 5 and 6 report the estimation results for the RRA for filtered and unfiltered log returns, respectively. *H* is the estimated Hurst exponent for the log returns series. $H_S$ is the estimated Hurst exponent for the time scales 5≤ n ≤ 40. $H_L$ is the estimated Hurst exponent for the time scales 40≤ n ≤ 299. For the Monte Carlo simulations, we report the mean estimated Hurst exponent over 5,000 replications, and the 0.005, 0.025, 0.005, 0.950, 0.975 and 0.995 quantiles.

Table 5 indicates that for the pre-filtered log returns series, the Hurst exponents estimated over the time scales $5 \leq n \leq 299$ are higher than the average *H* obtained from the 5,000 Monte Carlo simulations ($\mu_H$ = 0.572) in all cases but for the DJIA. However, none of these values exceeds the critical value at the 0.1 significance level, except for the PX-Glob, for which *H* exceeds the critical value at the 0.01 significance level. Accordingly, a two-tail test fails to reject the null hypothesis of *H* = 0.5 (log returns are temporally independent at all time scales) in favour of the alternative *H* ≠ 0.5 (long-range dependence at all time scales) for any of the indices except for the PX-Glob. Since the US stock market is supposed to be the most efficient, while the Czeck stock market should be relatively under-developed and inefficient, our results suggest that degree of long-range dependence is detected correctly.

For 5 ≤ *n* ≤ 40, the estimated Hurst exponent is smaller (for the CDAX and SWX), larger (for the FTSE 350, the ASE, and the PX-Glob), or equal (for the Mibtel and MSE) to the mean Hurst exponent obtained from Monte Carlo simulations ($\mu_H = 0.606$). However, as before, none of these values is higher than the critical value for the 0.1 significance level, except for the PX-Glob. Similarly, for $40 \leq n \leq 299$ the estimated Hurst exponent for the indices is neither smaller nor larger than the critical values for all indices, except for the PX-Glob, for which $H_L$ exceeds the critical value at the 0.05 significance level. However, the estimated Hurst exponent is larger than the mean Hurst exponent obtained from the Monte Carlo simulations ($\mu_H = 0.540$) for all indices, except for the DJIA.

[insert Table 5 here]

Table 6 reports the results for the unfiltered log returns series. The estimated Hurst exponent for the FTSE 350 is lower than $\mu_H$, and lower than the estimated Hurst exponent for the same series pre-filtered. In this case the pre-filtering eliminates significant negative autocorrelation for lags 3, 5, 6, and 10. It is interesting to note that in such a case, pre-filtering might *increase* the probability of spurious detection of long-range dependence, contrary to the received wisdom (see for example, Lo, 1991). For the Mibtel, the estimated $H$ is significantly larger than 0.5 at the 0.05 significance level. Accordingly, there is evidence of long-range dependence, which is not apparent when the Hurst exponent is estimated using the pre-filtered log returns series. Consistent with the results shown in table 5, the estimated $H$ for the PX-Glob is significantly larger than 0.5 at the 0.01 significance level. For $5 \leq n \leq 40$, for all indices except for the Mibtel, and the PX-Glob, $H_S$ for the unfiltered log returns series is lower than $H_S$ for the pre-filtered series, suggesting that pre-filtering may *increase* the chances of a spurious detection of long-range dependence. For the Mibtel and the PX-Glob, $H_S$ for the unfiltered series is higher than for the filtered series. For $40 \leq n \leq 299$, evidence of long-range dependence is obtained for the Mibtel and the PX-Glob. The estimated $H_L$ is different from 0.5 at the 0.1 significance level for the Mibtel and at the 0.01 significance level for the PX-Glob. Rejection of the null hypothesis for large time scales suggests that the detection of long-range dependence is not spurious, and is not driven by the presence of short-range dependence. For the FTSE 350, in common with the results for $5 \leq n \leq 299$, $H_L$ estimated for the unfiltered series is lower than H estimated for the pre-filtered series. As before, pre-filtering might increase rather than reduce the probability of detecting long-range dependence when it is not present.

[insert Table 6 here]

To summarise, while for the CDAX, the FTSE 350, the ASE, the MSE, the SWX, the PX-Glob and the DJIA the results for the filtered series are consistent with those for the unfiltered series, for the Mibtel the test for long-range dependence is affected by the pre-filtering procedure. This suggests that the appropriateness of pre-filtering is crucial for identifying long-range dependence for 'borderline' cases.

As shown in table 1, pre-filtering a series with both short-range and long-range dependence may erroneously dispose of the long-range dependence. In table 1 we consider the effect of pre-filtering using an AR(1) model on series with $H > 0.5$ (that is, $d > 0$). However, in tables 5 and 6 we have pre-filtered the series using the actual value of the PACF for lags for which the PACF is significant at the 0.05 level, rather than using an AR(1) model. Does pre-filtering lead to a large type II error even when the true values of the PACF are used for the pre-filtering?

We carry out an additional test to provide further insights on the existence of long-range dependence in the log returns of the Mibtel. Similar to section 2, we examine the impact of pre-filtering on the RRA estimate of the Hurst exponent when the true process exhibits short-term dependence only and when the true processes exhibits both long-range and short-range dependence. The PACF of the Mibtel is significant only for the 4$^{th}$ lag, that is, $\varphi_{44} = 0.0782$, where $\varphi_{ki}$ is the coefficient of the $i^{th}$ lag if the process has order $k$. We want to assess the effect of pre-filtering an ARFIMA($p$,$d$,0) series on the estimation of the fractional integration parameter $d = H - 0.5$, where $p = 4$, $\varphi_{44} = 0.0782$, and $\varphi_{43} = \varphi_{42} = \varphi_{41} = 0$. As before, we consider $T = 1000$ and $d = 0.00, 0.04, 0.08, 0.12$. The results (not reported but available upon request) show that for $d = 0.00$ the mean estimated Hurst exponent for the unfiltered and filtered series is $\hat{H} = 0.6281$ and $\hat{H} = 0.6132$, respectively. The critical values for the rejection of the null hypothesis of zero long-range dependence are 0.6385, 0.6459 and 0.6631 at the 0.1, 0.05 and 0.01 significance levels, respectively. Therefore, the presence of short-range dependence does not lead to a test that is over-sized. For $d = 0.04$, the mean estimated Hurst exponent is $\hat{H} = 0.6505$ for the unfiltered series and $\hat{H} = 0.6325$ for the filtered series. The mean estimated $H$ for the unfiltered series lies above the critical value at the 0.05 significance level, while the mean estimated $H$ for the filtered series lies below the critical value at the 0.1 significance level. Therefore, pre-filtering disposes of genuine long-range dependence for $d = 0.04$ and $\varphi_{44} = 0.0782$. For $d = 0.08$, the estimated Hurst exponent is $\hat{H} = 0.6729$ for the unfiltered series and $\hat{H} = 0.6509$ for the filtered series. In this case, the estimated $H$ for the unfiltered series lies above the critical value at the 0.01 significance level, and the estimated $H$ for the filtered series lies above the critical value at the 0.05 significance level. Therefore, pre-filtering does not lead to a particularly large type II error. For $d = 0.12$, the estimated Hurst exponent is $\hat{H} = 0.6952$ for the unfiltered series and $\hat{H} = 0.6680$ for the filtered series. In this case, both the estimated $H$ for the unfiltered series and the estimated $H$ for the filtered series lie above the critical value at the 0.01 significance level. The results for $d = 0.08$ and (even more) for $d = 0.12$ are consistent with our hypothesis that only for borderline cases is the pre-filtering able to dispose of genuine long-range dependence.

## 5  Conclusion

This paper presents an empirical analysis of the unifractal properties of the log returns of eight indices pertaining to stock markets that are believed to be in different stages of development and efficiency. Preliminary tests show evidence of non-Normality in the probability distribution function of the log returns calculated for various time scales. However, when temporal dependence is eliminated through a shuffle procedure, the distribution of the log returns calculated for large time scales tends towards Normality.

We test for evidence of long-range dependence in the log returns series for the eight market indices, using the Rescaled Range Analysis (RRA) to estimate the Hurst exponent. We employ the RRA because it is robust to non-Normality and it has been shown to be more powerful than alternative methods to detect long-range dependence. An estimated Hurst exponent significantly greater than 0.5 suggests long-range dependence in log returns.

Monte Carlo simulations are used to generate critical values for a test of the null hypothesis of no long-range dependence, taking into account the well-known upward bias in the RRA estimator of the Hurst exponent. We also use Monte Carlo simulations to examine the impact of pre-filtering for short-range dependence on the estimation of the Hurst exponent. Pre-filtering is a widely-employed technique to allow for short-term dependence.

We run the RRA on the log return series of the eight market indices and find evidence of long-range dependence, contrary to the EMH, for the Mibtel and the PX-Glob. The results for the other six indices do not reveal any evidence of long-range dependence. The results for the Mibtel become insignificant when the series is pre-filtered by an AR($p$) model. We show using Monte Carlo simulations that this method may reduce considerably the power of the test, while an examination of the series without pre-filtering does not necessarily result in a test that is over-sized.

Table 1: Descriptive statistics of estimated Hurst exponent for 1,000 Monte Carlo simulations of ARFIMA(1,d,0) processes with T = 1000.

| ρ = 0.20 | Mean H | SD(H) | Min(H) | Max(H) | ρ = 0.10 | Mean H | SD(H) | Min(H) | Max(H) |
|---|---|---|---|---|---|---|---|---|---|
| d=0.00 ρ=0.00 | 0.6130 | 0.0204 | 0.5534 | 0.6748 | d=0.00 ρ=0.00 | 0.6130 | 0.0204 | 0.5534 | 0.6748 |
| d=0.00 ρ=0.00* | 0.6129 | 0.0190 | 0.5537 | 0.6737 | d=0.00 ρ=0.00* | 0.6129 | 0.0190 | 0.5537 | 0.6737 |
| d=0.00 ρ=0.20 | 0.6654 | 0.0204 | 0.6079 | 0.7250 | d=0.00 ρ=0.10 | 0.6388 | 0.0204 | 0.5809 | 0.7000 |
| d=0.00 ρ=0.20* | 0.6130 | 0.0181 | 0.5535 | 0.6720 | d=0.00 ρ=0.10* | 0.6129 | 0.0186 | 0.5534 | 0.6731 |
| d=0.04 ρ=0.00 | 0.6348 | 0.0208 | 0.5749 | 0.6966 | d=0.04 ρ=0.00 | 0.6348 | 0.0208 | 0.5749 | 0.6966 |
| d=0.04 ρ=0.00* | 0.6240 | 0.0190 | 0.5618 | 0.6866 | d=0.04 ρ=0.00* | 0.6240 | 0.0190 | 0.5618 | 0.6866 |
| d=0.04 ρ=0.20 | 0.6854 | 0.0207 | 0.6265 | 0.7442 | d=0.04 ρ=0.10 | 0.6597 | 0.0208 | 0.6005 | 0.7204 |
| d=0.04 ρ=0.20* | 0.6203 | 0.0180 | 0.5587 | 0.6801 | d=0.04 ρ=0.10* | 0.6222 | 0.0185 | 0.5601 | 0.6838 |
| d=0.08 ρ=0.00 | 0.6566 | 0.0211 | 0.5954 | 0.7181 | d=0.08 ρ=0.00 | 0.6566 | 0.0211 | 0.5954 | 0.7181 |
| d=0.08 ρ=0.00* | 0.6339 | 0.0189 | 0.5689 | 0.6976 | d=0.08 ρ=0.00* | 0.6339 | 0.0189 | 0.5689 | 0.6976 |
| d=0.08 ρ=0.20 | 0.7053 | 0.0209 | 0.6436 | 0.7649 | d=0.08 ρ=0.10 | 0.6807 | 0.0211 | 0.6200 | 0.7407 |
| d=0.08 ρ=0.20* | 0.6260 | 0.0178 | 0.5627 | 0.6859 | d=0.08 ρ=0.10* | 0.6301 | 0.0184 | 0.5657 | 0.6922 |
| d=0.12 ρ=0.00 | 0.6785 | 0.0214 | 0.6154 | 0.7392 | d=0.12 ρ=0.00 | 0.6785 | 0.0214 | 0.6154 | 0.7392 |
| d=0.12 ρ=0.00* | 0.6423 | 0.0188 | 0.5749 | 0.7065 | d=0.12 ρ=0.00* | 0.6423 | 0.0188 | 0.5749 | 0.7065 |
| d=0.12 ρ=0.20 | 0.7249 | 0.0211 | 0.6609 | 0.7859 | d=0.12 ρ=0.10 | 0.7015 | 0.0213 | 0.6391 | 0.7617 |
| d=0.12 ρ=0.20* | 0.6300 | 0.0176 | 0.5654 | 0.6894 | d=0.12 ρ=0.10* | 0.6364 | 0.0183 | 0.5699 | 0.6985 |

Notes:
T is the number of observations of each simulated series. ρ is the first-order autocorrelation. d is the long-range dependence parameter.
* Indicates that the simulated series is pre-filtered by fitting an AR(1) model.

Table 2: Power of one-tail tests of $H_0: H=0.5$ against $H_1: H>0.5$, significance level = 0.05.

| Observations | 1,000 | 2,000 | 5,000 | 10,000 |
|---|---|---|---|---|
| **Rescaled Range Analysis** | | | | |
| H=0.54 | 0.290 | 0.448 | 0.717 | 0.876 |
| H=0.58 | 0.685 | 0.912 | 0.996 | 1.000 |
| H=0.62 | 0.931 | 0.996 | 1.000 | 1.000 |
| **Scaling function** | | | | |
| H=0.54 | 0.162 | 0.214 | 0.297 | 0.348 |
| H=0.58 | 0.378 | 0.496 | 0.675 | 0.79 |
| H=0.62 | 0.617 | 0.762 | 0.918 | 0.968 |

**Table 3: Descriptive statistics for the log return series calculated for the time scales: daily, weekly, monthly, and quarterly.**

|  | Mean | | | | Standard deviation | | | |
| --- | --- | --- | --- | --- | --- | --- | --- | --- |
|  | D | W | M | Q | D | W | M | Q |
| Mibtel | 0.0004 | 0.0018 | 0.0077 | 0.0230 | 0.0128 | 0.0285 | 0.0644 | 0.1037 |
| CDAX | 0.0002 | 0.0009 | 0.0041 | 0.0123 | 0.0138 | 0.0296 | 0.0654 | 0.1062 |
| FTSE350 | 0.0002 | 0.0008 | 0.0036 | 0.0109 | 0.0103 | 0.0212 | 0.0400 | 0.0661 |
| ASE | 0.0003 | 0.0013 | 0.0055 | 0.0165 | 0.0135 | 0.0284 | 0.0589 | 0.0982 |
| MSE | 0.0005 | 0.0024 | 0.0103 | 0.0310 | 0.0124 | 0.0268 | 0.0585 | 0.0957 |
| SWX | 0.0003 | 0.0015 | 0.0066 | 0.0199 | 0.0124 | 0.0263 | 0.0530 | 0.0898 |
| PX-GLOB | 0.0004 | 0.0022 | 0.0097 | 0.0290 | 0.0099 | 0.0253 | 0.0569 | 0.1049 |
| DJIA | 0.0003 | 0.0016 | 0.0068 | 0.0204 | 0.0110 | 0.0228 | 0.0463 | 0.0735 |
|  | Skewness | | | | Kurtosis | | | |
|  | D | W | M | Q | D | W | M | Q |
| Mibtel | -0.2069 | -0.3173 | 0.2048 | 0.4449 | 6.1328 | 5.0851 | 3.8485 | 3.2967 |
| CDAX | -0.2404 | -0.4544 | -0.8957 | -0.4392 | 5.9983 | 5.9294 | 5.5214 | 3.3556 |
| FTSE350 | -0.2426 | -0.3489 | -0.9070 | -0.8632 | 5.9985 | 5.0516 | 4.1338 | 4.5561 |
| ASE | -0.2144 | -0.6814 | -0.8963 | -1.0305 | 6.4857 | 5.6720 | 4.3286 | 4.2485 |
| MSE | -0.2481 | -0.5190 | -0.6175 | -0.4311 | 5.7933 | 5.5297 | 4.4316 | 3.1450 |
| SWX | -0.1546 | -0.3278 | -1.0693 | -0.8010 | 7.2722 | 6.9835 | 5.3010 | 4.0282 |
| PX-GLOB | -0.2910 | -0.5597 | 0.0284 | -0.3888 | 4.8890 | 3.9687 | 3.0830 | 2.9691 |
| DJIA | -0.2353 | -0.5457 | -0.6575 | -0.3976 | 7.1485 | 6.0364 | 4.1450 | 3.5428 |

Notes:

D = daily, W = weekly, M = monthly, and Q = quarterly. The number of observations for each sample frequency is: 2,608 (D), 522 (W), 120 (M) and 40 (Q).

**Table 4: Descriptive statistics for the shuffled log return series calculated for the time scales: daily, weekly, monthly, and quarterly.**

|  | Mean | | | | Standard deviation | | | |
|---|---|---|---|---|---|---|---|---|
|  | D | W | M | Q | D | W | M | Q |
| Mibtel | 0.0004 | 0.0018 | 0.0077 | 0.0230 | 0.0128 | 0.0284 | 0.0594 | 0.1053 |
| CDAX | 0.0002 | 0.0009 | 0.0041 | 0.0123 | 0.0138 | 0.0306 | 0.0576 | 0.1050 |
| FTSE350 | 0.0002 | 0.0008 | 0.0036 | 0.0109 | 0.0103 | 0.0233 | 0.0479 | 0.0719 |
| ASE | 0.0003 | 0.0013 | 0.0055 | 0.0165 | 0.0135 | 0.0296 | 0.0678 | 0.1189 |
| MSE | 0.0005 | 0.0024 | 0.0103 | 0.0310 | 0.0124 | 0.0277 | 0.0568 | 0.1011 |
| SWX | 0.0003 | 0.0015 | 0.0066 | 0.0199 | 0.0124 | 0.0247 | 0.0529 | 0.1008 |
| PX-GLOB | 0.0004 | 0.0022 | 0.0097 | 0.0290 | 0.0099 | 0.0220 | 0.0441 | 0.0782 |
| DJIA | 0.0003 | 0.0016 | 0.0068 | 0.0204 | 0.0110 | 0.0245 | 0.0515 | 0.0871 |

|  | Skewness | | | | Kurtosis | | | |
|---|---|---|---|---|---|---|---|---|
|  | D | W | M | Q | D | W | M | Q |
| Mibtel | -0.2069 | 0.0058 | -0.1245 | 0.2896 | 6.1328 | 3.5781 | 3.3671 | 2.8103 |
| CDAX | -0.2404 | -0.2658 | 0.0869 | -0.1896 | 5.9983 | 3.3644 | 3.6264 | 2.8605 |
| FTSE350 | -0.2426 | -0.1545 | -0.3739 | -0.3915 | 5.9985 | 3.5712 | 2.7612 | 3.1666 |
| ASE | -0.2144 | 0.0029 | 0.3306 | 0.3772 | 6.4857 | 3.1832 | 3.6534 | 2.6888 |
| MSE | -0.2481 | -0.1944 | -0.3888 | -0.2709 | 5.7933 | 3.6475 | 3.6577 | 2.2852 |
| SWX | -0.1546 | -0.2709 | 0.1806 | -0.2209 | 7.2722 | 3.5318 | 2.9351 | 2.6644 |
| PX-GLOB | -0.2910 | -0.3340 | -0.1610 | -0.0310 | 4.8890 | 3.3037 | 3.0129 | 1.8668 |
| DJIA | -0.2353 | -0.1083 | 0.1085 | 0.3969 | 7.1485 | 3.3997 | 3.1377 | 3.7133 |

Notes:

D = daily, W = weekly, M = monthly, and Q = quarterly. The number of observations for each sample frequency is: 2,608 (D), 522 (W), 120 (M) and 40 (Q).

**Table 5: Rescaled Range Analysis results (with pre-filtering).**

| Index | H | $H_s$ | $H_L$ | | Monte Carlo simulations | | |
|---|---|---|---|---|---|---|---|
| | | | | | H | $H_s$ | $H_L$ |
| Mibtel | 0.593 | 0.606 | 0.580 | $\mu_H$ | 0.572 | 0.606 | 0.540 |
| CDAX | 0.592 | 0.591 | 0.593 | | | | |
| FTSE350 | 0.585 | 0.617 | 0.554 | | *Quantiles* | | |
| ASE | 0.596 | 0.614 | 0.578 | 0.005 | 0.528 | 0.562 | 0.451 |
| MSE | 0.590 | 0.606 | 0.573 | 0.025 | 0.539 | 0.573 | 0.475 |
| SWX | 0.598 | 0.603 | 0.593 | 0.050 | 0.545 | 0.579 | 0.484 |
| PX-GLOB | 0.631*** | 0.638* | 0.624** | 0.950 | 0.601 | 0.633 | 0.597 |
| DJIA[+] | 0.568 | 0.601 | 0.535 | 0.999 | 0.618 | 0.650 | 0.630 |

Notes:
$H$ is the estimated Hurst exponent for the log returns series. $H_S$ is the estimated Hurst exponent for the time scales $5 \leq n \leq 40$. $H_L$ is the estimated Hurst exponent for the time scales $40 \leq n \leq 299$. The log returns series are pre-filtered using an AR($p$) model where $p$ is the lags for which the PACF is significant at the 5% level. The lags used for each index are: Mibtel: 4th lag; CDAX: 6th and 8th lag; FTSE 350: 3rd, 5th, 6th, 8th, 10th lags; ASE: 3rd, 5th, 8th, 9th lags; MSE: 3rd, 8th, 10th lags; SWX: 5th lag; PX-GLOB: 1st and 4th lag. For the Monte Carlo simulations, we report the mean estimated Hurst exponent and the 0.005, 0.025, 0.005, 0.950, 0.975 and 0.995 quantiles.
[+] For the DJIA the Autocorrelation Function for the lags from 1 to 10 does not suggests significant short-range dependence. The results reported in tables 5 and 6 are the same.
\*\*\*, \*\*, \* denote rejection of the null hypothesis of no long-range dependence at the 0.01, 0.05 and 0.1 significance levels, respectively.

**Table 6: Rescaled Range Analysis (without pre-filtering).**

| Index | $H$ | $H_s$ | $H_L$ | | Monte Carlo simulations | | |
| --- | --- | --- | --- | --- | --- | --- | --- |
| | | | | | $H$ | $H_s$ | $H_L$ |
| Mibtel | 0.613** | 0.623 | 0.602* | $\mu_H$ | 0.572 | 0.606 | 0.540 |
| CDAX | 0.588 | 0.586 | 0.590 | | | | |
| FTSE350 | 0.553 | 0.586 | 0.520 | | *Quantiles* | | |
| ASE | 0.588 | 0.588 | 0.587 | 0.005 | 0.528 | 0.562 | 0.451 |
| MSE | 0.595 | 0.593 | 0.597 | 0.025 | 0.539 | 0.573 | 0.475 |
| SWX | 0.585 | 0.594 | 0.577 | 0.050 | 0.545 | 0.579 | 0.484 |
| PX-GLOB | 0.665*** | 0.676*** | 0.655*** | 0.950 | 0.601 | 0.633 | 0.597 |
| DJIA | 0.568 | 0.601 | 0.535 | 0.999 | 0.618 | 0.650 | 0.630 |

Notes:
$H$ is the estimated Hurst exponent for the log returns series. $H_S$ is the estimated Hurst exponent for the time scales $5 \leq n \leq 40$. $H_L$ is the estimated Hurst exponent for the time scales $40 \leq n \leq 299$. The log returns series are unfiltered. For the Monte Carlo simulations, we report the mean estimated Hurst exponent and the 0.005, 0.025, 0.005, 0.950, 0.975 and 0.995 quantiles.

**, * denote rejection of the null hypothesis of no long-range dependence at the 0.05 and 0.1 significance levels, respectively.